# Loophole-free Einstein-Podolsky-Rosen experiment via quantum steering

Bernhard Wittmann*[1,2], Sven Ramelow*[1,2], Fabian Steinlechner[2], Nathan K. Langford[2], Nicolas Brunner[3], Howard M. Wiseman[4], Rupert Ursin[2] and Anton Zeilinger[1,2]

[1] Vienna Center for Quantum Science and Technology (VCQ), Faculty of Physics,
University of Vienna, Boltzmanngasse 5, A-1090, Vienna, Austria
[2] Institute for Quantum Optics and Quantum Information (IQOQI),
Austrian Academy of Sciences, Boltzmanngasse 3, A-1090, Vienna, Austria
[3] H.H. Wills Physics Laboratory, University of Bristol, Bristol BS8 1TL, United Kingdom
[4] Centre for Quantum Computation and Communication Technology (Australian Research Council), Centre for Quantum
Dynamics, Griffith University, Brisbane, Queensland 4111, Australia

*these authors contributed equally to this work

Tests of the predictions of quantum mechanics for entangled systems have provided increasing evidence against local realistic theories. However, there still remains the crucial challenge of simultaneously closing all major loopholes – the locality, freedom-of-choice, and detection loopholes – in a single experiment. An important sub-class of local realistic theories can be tested with the concept of "steering". The term steering was introduced by Schrödinger in 1935 for the fact that entanglement would seem to allow an experimenter to remotely steer the state of a distant system as in the Einstein-Podolsky-Rosen (EPR) argument. Einstein called this "spooky action at a distance". EPR-Steering has recently been rigorously formulated as a quantum information task opening it up to new experimental tests. Here, we present the first loophole-free demonstration of EPR-steering by violating three-setting quadratic steering inequality, tested with polarization entangled photons shared between two distant laboratories. Our experiment demonstrates this effect while simultaneously closing all loopholes: both the locality loophole and a specific form of the freedom-of-choice loophole are closed by having a large separation of the parties and using fast quantum random number generators, and the fair-sampling loophole is closed by having high overall detection efficiency. Thereby, we exclude – for the first time loophole-free – an important class of local realistic theories considered by EPR. As well as its foundational importance, loop-hole-free steering also allows the distribution of quantum entanglement secure from an untrusted party.

1.  Introduction

According to quantum theory, when two systems are "entangled", a local measurement performed on one of them instantaneously collapses the state of the other distant one. Einstein called this "spooky action at a distance"[1] and argued that the quantum state cannot describe the "real factual situation", because it depends on the type of measurement performed on a distant system[2]. In their famous 1935 paper Einstein, Podolsky, and Rosen (EPR)[3] used this effect to show that there is a deep conflict between the quantum formalism and the principle that a spatially isolated system should be completely described by local properties. Quantum mechanics seems to predict the ability to instantaneously influence a remote quantum state at arbitary distances. Schrödinger gave the name "steering" to the possibility of remotely piloting a state, more than any classical correlations would allow, by pointing out: "It is rather discomforting that the theory should allow a system to be steered or piloted into one or the other type of state at the experimenter's mercy in spite of his having no access to it."[4] It is only recently that this "steering" has been rigorously formulated[5,6], allowing the derivation of an EPR-steering inequality[7] from the assumption that the remote system can be described by local quantum mechanics only. If such an inequality is violated experimentally, this demonstrates EPR steering. This was implemented recently by Saunders *et al* using polarization-entangled photons, although without closing any loopholes[8].

By violating a three-setting steering inequality using polarization entangled photons, shared between two distant observers we demonstrate EPR-steering in a loophole-free fashion. This is done by realizing space-like separation of all relevant events to close the locality loophole[9-11] and a specific form of the freedom-of-choice loophole[12,13] (explained in detail in the experimental section) and by simultaneously detecting a large enough sub-ensemble to close the fair-sampling loophole[14-18]. Thus our experiment provides, for the first time, a loophole-free test of quantum steering using entanglement.

In an EPR-steering experiment (see Fig. 1 top), Alice delivers a state to Bob, who only trusts in local quantum mechanics. Alice claims to be able to remotely steer Bob's state, but he is skeptical and requires Alice to prove this.

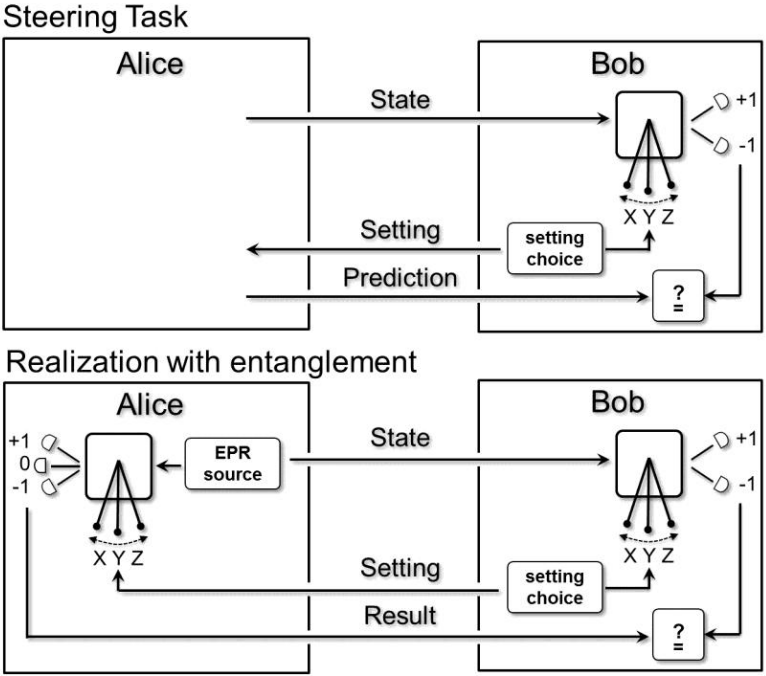

**Figure 1: Top:** In a steering experiment, Alice sends a system to Bob that he assumes to be an unknown local quantum state[5-7]. Next Bob chooses freely in which setting (X, Y or Z) to measure. Then he sends his choice of setting to Alice and records secretly his measurement result. Bob now challenges Alice, who claims that she can steer his state from a distance, to predict his result (+1, -1). Provided the correlation between her prediction and his result is above the steering bound, Bob is forced to conclude Alice indeed remotely steered his state (spooky action at a distance), or give up his assumption of a local quantum state. **Bottom:** Using entangled pairs of photons produced by an EPR source Alice can demonstrate steering. She measures her photon with the same setting Bob announced. Entanglement ensures (anti-)correlations between Alice's and Bob's outcomes for all measurement choices and allows to violate the steering bound. To close the fair sampling loophole one must also account for Alice's inconclusive (0) results when she detects no photon and include these results when calculating the steering value.

Upon receiving a state from Alice, he chooses a measurement setting and announces this to Alice. He then challenges Alice to predict the result of his measurement (which he keeps secret). Bob can work out how well-correlated Alice's prediction can be with his outcome, given the assumption that he holds a local quantum state[5]. Bob's local quantum state is represented by a density matrix, which can be unknown to him, though perhaps known to Alice. It is local in the sense that Alice cannot influence it because she has no physical access to it. The bound on Alice's and Bob's correlation under this assumption is known as the EPR-steering bound. If the correlation Bob measures (after Alice has announced her guess) is above the EPR-steering bound, Bob must reject the local quantum state assumption.

2. **Theory**

In our steering experiment Alice sends Bob one photon of an entangled two-photon state (see Fig. 1, bottom). As with experimental tests of Bell's theorem, imperfections in the experimental implementation can potentially open up "loopholes", which would allow apparent violations of the inequalities to be explained by a model that still fulfills the basic assumptions about locality. The three major loopholes are based on: hidden communication between the observers (locality loophole)[9-11], possible influences from or on the choice of measurement settings (freedom-of-choice loophole)[12,13,] and unfair sampling of the measured ensemble (detection loophole)[14-18]. In our experiment - for the first time in any experiment - all of these loopholes are closed simultaneously.

We test a steering inequality using three measurement settings for both Alice and Bob, which are set to be the mutually unbiased qubit bases X, Y, and Z. For each basis, the measurement result is thus a binary variable, +1 or -1. However, because of the limited transmission and detection efficiency of the photons, a third outcome must also be considered, 0, which corresponds to no detection, giving in total three possible (ternary) measurement outcomes $r \in \{+1, -1, 0\}$. Bob tests, and therefore trusts, his measurement apparatus, including his detectors, and is free to choose which of his measurement events Alice has to predict, so inconclusive events (0) can be discarded on his side. However, because Bob does not trust Alice, he does not allow her to discard any results; if Alice claims to obtain inconclusive results then her output must be treated as truly ternary $r \in \{+1, -1, 0\}$. In each run of the experiment, Bob chooses one of the three possible measurement settings and records his measurement outcome as well as Alice's prediction. In order to make sure that the total statistics are incompatible with any model that assumes a local quantum state, Bob then tests the following steering inequality [5]:

$$S = T_X + T_Y + T_Z \leq 1 \qquad (1)$$

where

$$T_X = \sum_{r=+1,-1,0} P(r|X_A)[< X_B >|_{X_A=r}]^2 \qquad (2)$$

and $T_Y$, $T_Z$ are defined similarly, $P(r|X_A)$ is the probability that Alice obtains the result r given the setting X, and $< X_B >|_{X_A=r}$ is Bob's average for the (binary) outcome of his measurement of the qubit operator X from that subensemble where Alice reports r. Thus $T_X$, $T_Y$ and $T_Z$ represent the respective correlations between Alice's and Bob's outcomes for the three bases. The fact that inequality (1) holds for any model that assumes a local quantum state for Bob, follows from the fact that any state is represented by a vector within a unit Bloch sphere, which satisfies

$$< X_B >^2 + < Y_B >^2 + < Z_B >^2 \leq 1 \qquad (3)$$

Each term on the left-hand-side here is a convex function of the quantum state: the value of $< X_B >^2$ (for example) for a state which is a weighted mixture of any ensemble of states is less than or equal to the weighted average of $<X_B>^2$ across this ensemble. Now if Bob does have a local quantum state, chosen from some ensemble by Alice (and hence known to her), $[<X_B>|_{X_A=r}]^2$ equals the square of the average of $<X_B>$ in the subensemble where Alice, if asked to predict $X_B$, would give the prediction $X_{A=r}$. That is, it equals the square of the expectation value of $X_B$ for the average state from the subensemble in which Alice would predict $X_{A=r}$. But by the convexity property, this is bounded above by the ensemble average of $<X_B>^2$ across that subensemble. Now, averaging over all three values of r to obtain the full ensemble average $T_X$ as in Eq. (2), and adding this together with $T_Y$ and $T_Z$ (which each corresponds to the same full ensemble) are all linear operations. Thus the final result is still bounded by the ensemble average of $<X_B>^2+<Y_B>^2+<Z_B>^2$ across the full ensemble. But since every member of the ensemble obeys Eq. (3), we thus obtain the local quantum state bound in Eq. (1).

Uncertainty relations are central in the study of the foundations of quantum mechanics. Therefore it is interesting to note that inequality (3) is equivalently fulfilled for any local realistic model describing Bob's system, with the additional assumption of the quadratic uncertainty relation on his side only. Note that, such an uncertainty relation can be fulfilled by probabilistic local realistic theories. Therefore, the class of theories that is tested by the steering inequality (1) consists of probabilistic local realistic theories that fulfill a quadratic uncertainty relation on one side. Whether all such probabilistic theories can be represented by ensembles of deterministic theories we leave open.

Inequality (1) can be violated by a large class of entangled states. For example, if Alice and Bob share a singlet state $\psi^- = \frac{1}{\sqrt{2}}(|10> - |01>)$, and Alice has perfect detectors, she can achieve the maximal violation of S = 3, because identical measurements on the singlet state always lead to perfectly anti-correlated outcomes[19]. Most importantly, since inequality (1) takes Alice's inconclusive events into account, it does not rely on a fair-sampling assumption. Hence, any experiment able to violate this inequality closes the detection loophole. If the locality and the freedom-of-choice loopholes are also closed, then an experimental violation of this inequality can be considered loophole-free.

### 3. Setup

Bob's lavatory (see small laboratory in the floor plan of Fig. 2) is spatially separated from Alice's laboratory by 48 m of direct distance and connected via a single-mode fibre quantum channel and a classical link. Photon pairs are produced by a fibre-coupled source based on type-II spontaneous parametric down-conversion via a non-linear crystal in a Sagnac loop at 810 nm, pumped by a violet (405 nm) laser[20,21]. One photon of each pair is kept locally, and its partner photon, sent via the quantum channel to Bob, is the quantum state Alice is challenged to steer. High arm efficiency (coupling, transmission and detection) on Alice's side was achieved by optimizing focusing parameters in the source, fully suppressing counts from the pump laser by using cut-off filters in both arms, and using a 0.5 nm interference filter on Bob's side. On Bob's side, his home-made quantum random number generators (QRNGs) produce two random bits (00, 10 , 01 or 11) to select one of three orthogonal settings X, Y or Z (ignoring the last combination) which he sends to Alice via the classical link. These settings correspond to the polarization bases +45°/-45°, R/L or H/V, respectively where H (V) denotes horizontal (vertical) polarization, +45° (-45°) diagonal (anti-diagonal) linear polarization, and R (L) right-hand (left-hand) circular polarization. Alice performs the corresponding polarization measurement using two electro-optical modulators (EOMs) and two single-photon detectors monitoring the outputs of a polarization beam splitter (PBS). Alice's result (the prediction which she makes for Bob's result) is immediately sent back to Bob via coaxial cables. If Alice detects no photon, Bob counts this as an inconclusive event from Alice (0). An identical module is located on Bob's side measuring his photon using the same setting.

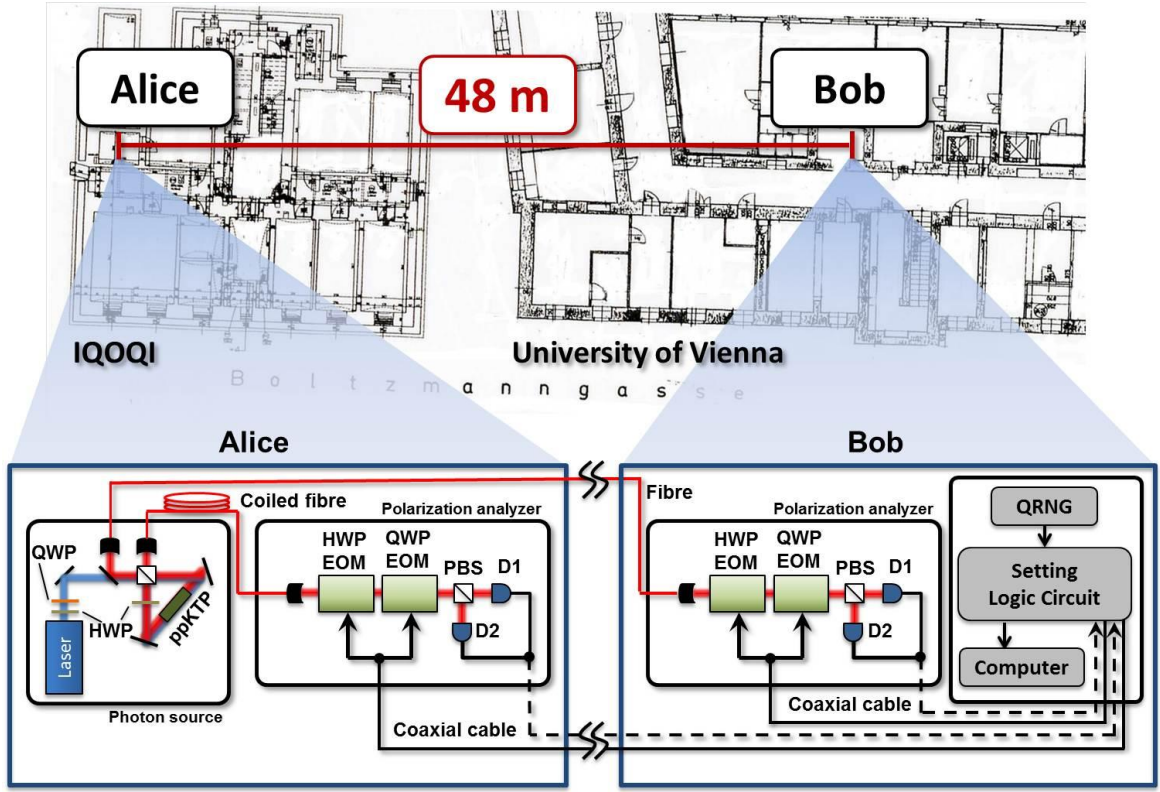

**Figure 2: Experimental setup.** The loophole-free steering experiment was carried out between two buildings: the University of Vienna and the Austrian Academy of Sciences (IQOQI). A polarization-entangled photon pair is generated by Alice using an entangled photon source[20,21]. For each entangled pair, one photon is kept in an 80 m long, coiled optical single-mode fibre (red line) on Alice's side, located next to the source. The other photon is sent to Bob via another optical fibre. On Bob's side, one of three measurement settings is chosen by his fast home-made quantum random number generators (QRNG) based on [22] and sent to Alice's side via the classical link. The setting choice is stored locally and also sent to Alice via a low-dispersion coaxial link and subsequently applied to both photons (solid black lines). Alice's polarization analyzer implements the different settings with two electro-optical modulators (EOMs) realizing ultra-fast switchable half- and quarter-wave plates (HWP, QWP), as well as a polarizing beam splitter (PBS) and two home-made photon detector modules based on silicon avalanche photo-diodes ($D_i$). On Bob's side there is an equivalent polarization analyzer. The results are then collected (dashed black lines) in Bob's lab and compared in a logic circuit.

Coincident detection events identify the two photons of the distributed photon pair and are registered by a fast electronic AND gate, implemented on a field programmable gate array (FPGA) board. Both the coincidences and the single counts of both parties are recorded together with the measurement settings by a computer. The steering parameter is then directly calculated from the measured data according to equation (1), without any background noise subtraction.

## 4. Space-Time Arrangement.

In our experiment, the locality loophole would arise if Alice were able to exploit information about Bob's setting choice or measurement outcome by any luminal or sub-luminal signal. The freedom-of-choice loophole would imply the possibility of a causal connection between the setting choice and the photon-pair emission.

In particular, we close here a specific form of the freedom of choice loophole, namely the loophole which would allow the photon pair emission to influence the quantum random number generator that chooses the measuring set up. What we cannot exclude, as with any experiment, is the possibility that an earlier common cause in the overlap of the backward light cones of the two events (emission and choice of the setting), influences the two events in a correlated manner. We believe, however, that such a hypothesis is outside the scope of what can in principle be tested experimentally[12,25]. We simultaneously close these two loopholes (locality and freedom-of-choice) by fulfilling several critical conditions relating to the space-time arrangement of the relevant events. They are illustrated and explained in detail in Figure 3 and its caption.

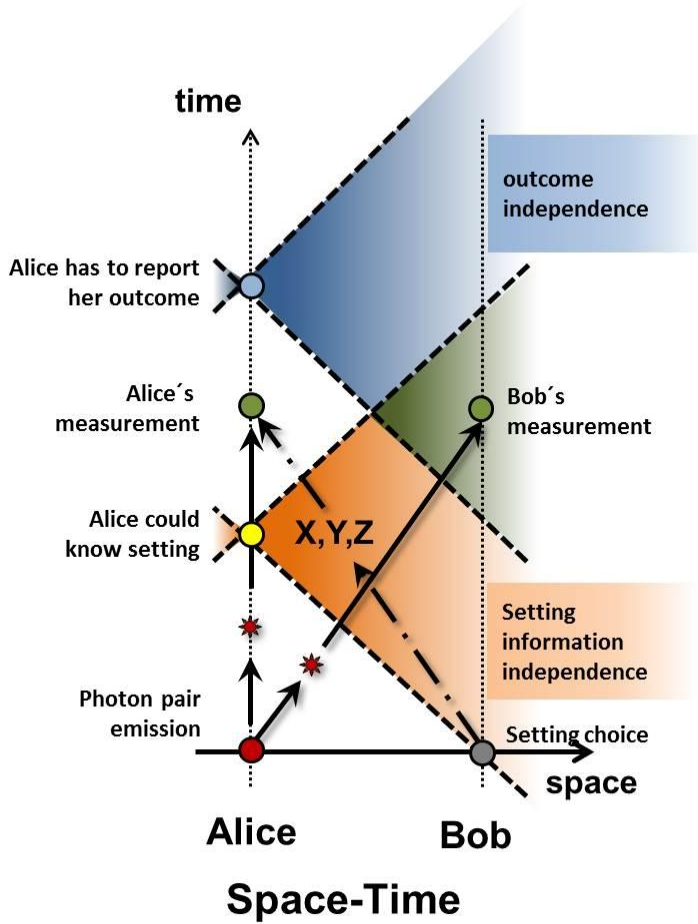

**Figure 3: Space-time diagram illustrating the conditions to close the locality and freedom-of-choice loopholes (illustration, not to scale):** Firstly, Alice creates a photon pair and sends a photon to Bob before she receives Bob's setting choice. In our experiment, we enforce freedom of choice by space-like separating the generation of the entangled photon pairs (red dot), and the QRNG´s choice (grey dot). They happen 48 m apart at the same time (t=0) in the lab frame. The second condition is to exclude any causal influence of Alice on Bob's measurement once she can know the setting. This setting information independence is enforced in our experiment by Bob's measurement taking place in a region (orange area) that is space-like separated from the event (yellow dot) marking the time at which Alice could in principle know Bob's setting. Lastly, we also ensure that Alice cannot know Bob's measurement outcome before reporting hers, since reproducing any arbitrary correlation would then become a trivial task for her. This outcome independence is guaranteed by the event of Alice's outcome report (blue dot) being space-like separated (blue area) from Bob´s measurement event (left green dot). Timing Bob's measurement such that setting information independence and outcome independence are enforced simultaneously (green area) closes the locality loophole. For simplicity the different events are illustrated with a dot, not with the actual time they need in the experiment.

The time window of the setting information independence is determined by the distance between the labs - 360 ns. To calculate the overlap of this time window and the space-time region where simultaneously outcome independence is enforced we have to take the production of the random number, all transfer, switching and measurement times into account: our measurements are triggered by an external clock (t = 0) with a rate of 787 kHz on Bob's side, sampling the outputs of two home-made QRNG's based on [22] at t = 90 ns. This time interval takes into account an internal electronic delay of 45 ns, as well as three autocorrelation times to assure that no information about the QRNG's choice is present before t = 0. The transmission of the setting produced in the QRNG at t = 90 ns then takes 205 ns. Afterwards the random setting is applied by using EOMs with a switching time of 22 ns. In front of them is a splitter box to convert the previously amplified TTL signal into a useful signal for the EOMs which in total takes (including all cables) 48 ns. In addition, we delay the measurement time on Alice side by 20 ns to find the best visibility for the measurement. It then takes a few hundred picoseconds from the photon impact on the detector until the Si-APD breakdown and avalanche come to a halt and less than 10 ns until the electronic signal is registered[23]. From that point on we regard the detection event as completed – we assume that such a classical signal (click) is immune to modification by any hypothetical influence. The spatial separation and all the transmission, switching and measurement times lead us to a trusted time window of 75 ns (green area in Fig. 2). To simultaneously guarantee setting information independence and outcome independence, Bob only considers measurement results during 20 ns (including the registration time in the detector), placed in the middle of this trusted window with a 25 ns buffer both to the beginning and the end of the trusted area. Importantly, also the production of the photon pairs and the full region of where the random choice is made (90 ns) by our QRNG are space-like separated, enforcing their causal independence.

5. Results

When the bound of the steering inequality is surpassed (i.e. when S > 1 for mutually unbiased measurement settings; see Eq. 1), then steering is confirmed. Theoretically, quantum mechanics predicts, for a maximally entangled singlet state, a maximal violation of the steering inequality, i.e. $S_{th}$ = 3. In the experiment, however, there are number of factors which will reduce the measured steering parameter, including overall arm (coupling, transmission and detection) efficiency of Alice ($\eta$) and overall visibility (V). The steering value expected to be observed in the experiment is then given by $S = S_{th}\eta V^2$.

We measured an average total arm efficiency on Alice's side in the setup of 38.3±0.1%. This is well above the required minimum efficiency for loophole-free steering of 1/3 using inequality (1). The factors that lead to this efficiency are: the total arm efficiency of our source of 50.6%±0.1, which includes the efficiency of our detectors, optical losses and fibre coupling losses in the source. There is also a ~9% loss in Alice's 80 m delay fibre and ~16% loss in the polarization analysis module with two EOMs. Our overall visibility is reduced by the imperfect entanglement visibility of our source (around 98% in X-basis), combined with the non-ideal visibility of both our polarization analyzer modules (average around 98%) and the long-term stability of the fibre quantum channels (99.5%) over the measurement time of several hours without any active polarization stabilization. All these effects lead to a significant reduction of the experimentally observable steering parameter from its ideal quantum value of 3, with the detection efficiency for Alice's two APDs of around 60% having by far the biggest contribution.

If Bob's measurement settings are not perfectly mutually unbiased, Alice could choose a specific local quantum state that could yield a higher steering value than 1. We therefore carefully characterized Bob's polarization analyzer module with a novel form of measurement tomography (related to [24]). Measuring the response for a complete set of polarization states (H, V, +45°, -45°, R, L), we independently reconstruct the 6 different measurement operators describing our analyzer module using maximum likelihood optimization. They are very close to the ideal, slightly impure but almost perfectly mutually unbiased. We then calculate the highest steering value for any pure state (and thus by convexity for any local quantum state) Alice could achieve as 0.990±0.001. The error margin was determined by Monte-Carlo simulations based on Poissonian count statistics. Therefore, Alice in fact cannot reach a value higher than 1 and we – more conservatively – choose to compare our results against the ideal bound of 1.

We performed 360 runs, integrating the singles and coincidence counts for 30 seconds each. This resulted in a steering value of $S_{exp} = 1.049\pm0.002$, clearly violating the steering bound of 1 by more than 20 standard deviations. Moreover, in each basis we achieved a polarization correlation coefficient $T_i^{exp} > 1/3$. Three standard deviations were reached after less than 300 seconds measurement. The error is given by the standard deviation of the mean for the 360 measurements, and agrees very well with what one would expect from Poissonian count statistics. The detailed results for the different measurement bases are shown in Table 1. We emphasize that no kind of background noise subtraction was used to obtain these results. Since the measured steering value is above the bound, Bob is forced to conclude that Alice successfully steered his stat

|  | H/V Basis | +/- Basis | R/L Basis |
|---|---|---|---|
| Alice's arm efficiency$^{exp}$ | 38.2% ± 0.1% | 38.3% ± 0.1% | 38.3% ±0.1% |
| Total visibility$^{exp}$ | 96.23% ± 0.05% | 95.41% ± 0.06% | 95.05% ± 0.06% |
| $T_i^{exp}$ value | 0.354 ±0.001 | 0.349±0.001 | 0.347 ±0.001 |
| Steering value $S^{exp}$ | | 1.049 ± 0.002 | |

**Table 1 | Experimental results.**
We measured the polarization correlation coefficients $T_i$ closing simultaneously the locality and freedom-of-choice loopholes, to test the steering inequality (1) without a fair-sampling assumption. We obtained a measured steering value of $S_{exp}= 1.049\pm0.002$. All error margins were derived directly from the statistics of our count rates (standard deviation of the mean). The final steering value excludes any local hidden state model by more than 20 standard deviations. The observed steering value matches excellently what one expects from the measured overall arm efficiency of Alice ($\eta = 38.3\%$) and overall visibility (V = 95.56%) with $S = S_{th}\, \eta\, V^2$ and $S_{th} = 3$.

## 6. Discussion

Here, we demonstrate an experimental violation of our steering inequality using entangled photon pairs distributed over 48 m. The inequality takes into account null (0) results by Alice and so does not require any fair-sampling assumption, this is the first time this type of loophole has been closed in an experiment with photons. In addition, our experiment is realized under strict Einstein locality conditions to also close the locality and a specific form of the freedom-of-choice loopholes. Simultaneously closing these three major loopholes in a single experiment excluding an important sub-class of local realistic theories is a major step forward, particularly with regard to future loophole-free experiments testing Bell inequalities[25-29], which would exclude all local realistic theories. Beside the distribution of quantum entanglement to establish security from an untrusted party[30-32], loophole-free steering bears foundational importance because it demonstrates a non-local quantum effect for the first time in a loophole-free fashion. Our results show most rigorously that if one demanded that an isolated system is defined by a local quantum state, this would imply the existence of "spooky action at a distance".

**Acknowledgements**

The authors thank Caslav Brukner and Johannes Kofler for their stimulating discussions and Marissa Giustina and Harald Rossman for technical support. This work was supported by the FWF Doctoral Programme CoQuS (W 1210) and SFB FoQus, Austrian Research Promotion Agency (FFG), the European Comission under the Q-ESSENCE contract (248095), as well as by the Australian Research Council Centre of Excellence for Quantum Computation and Communication Technology (Project number CE110001027) and the UK EPSRC.

**Author Information**

Correspondence and requests for materials should be addressed to bernhard.wittmann@univie.ac.at and sven.ramelow@univie.ac.at